\documentstyle[12pt]{article}

\makeatletter
   \def\@cite#1#2{\leavevmode\hbox{$^{\mbox{\the\scriptfont0 #1}}$}}
\makeatother

     \def\sN{\scriptscriptstyle N}
     \def\sF{\scriptscriptstyle F}
     \def\ps{\hbox{\raisebox{.2ex}{$p$}}\mbox{\hspace{-0.9ex}}/}

     \def\nab{\nabla\mbox{\hspace{-1.5ex}}/}
     \def\yy{y\mbox{\hspace{-0.9ex}}y}
     \def\11{1\mbox{\hspace{-0.9ex}}1}

\makeatletter
\@addtoreset{equation}{subsection}

\makeatother

\makeatletter
\long\def\@makecaption#1#2{%
   \vskip 10\p@
   \setbox\@tempboxa\hbox{#1\ \ #2}%
   \ifdim \wd\@tempboxa >\hsize
   #1\ \ #2\par        %
      \else
   \hbox to\hsize{\hfil\box\@tempboxa\hfil}%
   \fi}
\def\fnum@figure{Fig. \thefigure}
\makeatother

\begin{document}

\begin{flushright}
\begin{tabular}{l}
HUPD-9516 \hspace{1em}\\
July 1995
\end{tabular}
\end{flushright}

\vspace{1.5cm}

\begin{center}
{\Large \bf
Scalar and Spinor Two-Point Functions\\[4mm]
in Einstein Universe}\\[2cm]
\normalsize
T.~Inagaki
\footnote{e-mail : inagaki@theo.phys.sci.hiroshima-u.ac.jp},
K.~Ishikawa
\footnote{e-mail : ishikawa@theo.phys.sci.hiroshima-u.ac.jp}
and T.~Muta
\footnote{e-mail : muta@sci.hiroshima-u.ac.jp}
\\[1cm]
{\it
Department of Physics, Hiroshima University, \\
Higashi-Hiroshima, Hiroshima 739, Japan \\[3cm]
}
\end{center}

\begin{abstract}
\vglue 0.7cm
Two-point functions for scalar and spinor fields are investigated
in Einstein universe ($R \otimes S^{\sN-1}$).
Equations for massive scalar and spinor two-point functions
are solved and the explicit expressions for the two-point
functions are given.
The simpler expressions for massless cases are obtained
both for the scalar and spinor cases.
\end{abstract}

\newpage


\renewcommand{\arraystretch}{2}
\renewcommand{\thesubsection}{\arabic{subsection}}
\renewcommand{\thesubsubsection}
   {\arabic{subsection}.\arabic{subsubsection}}
\baselineskip=23pt

It is important to investigate quantum field theories in curved space-time
in connection with phenomena under strong gravity as in the early
universe.
The fundamental object in dealing with quantum field theories
in curved space-time is the two-point Green function which we shall
study in the present communication.
There are numerous previous studies which have dealt with the
two-point functions in the maximally symmetric space-time
$S^{\sN}$ and/or $H^{\sN}$.
In the present paper we push forward the investigation and study
the Einstein universe.

We consider the manifold $R \otimes S^{\sN-1}$ as an Euclidean analog
of the $N$-dimensional Einstein universe.\cite{DA}
The manifold is defined by the metric
\begin{equation}
     ds^{2}=dr^{2}+a^{2}(d\theta^{2}+\sin^{2}\theta d\Omega_{\sN-2})\, ,
\end{equation}
where $d\Omega_{\sN-2}$ is the metric on a unit sphere $S^{\sN-2}$.
The manifold is a constant curvature space with curvature
\begin{equation}
     R=(N-1)(N-2)\frac{1}{a^{2}}\, .
\end{equation}

We start with the argument on scalar two-point functions.
On the manifold $R \otimes S^{\sN-1}$ the scalar two-point function is
defined by the equation
\begin{equation}
     ((\partial_{0})^{2}+\Box_{\sN-1}-\xi R-m^{2})G_{\sF}(y_{(0)},y)
     =-\frac{1}{\sqrt{g}}\delta^{\sN}(y_{(0)},y)\, ,
\label{eq:GF}
\end{equation}
where $g$ is the determinant of the metric tensor $g_{\mu\nu}$,
$\Box_{\sN-1}$ is the Laplacian on $S^{\sN-1}$,
$\delta^{\sN}(y_{(0)},y)$ the Dirac delta function in the manifold
$R \otimes S^{\sN-1}$ and
$\xi$ the coupling constant between the scalar field and the curvature.
In the following discussions we fix $y_{(0)}$ at the origin
and write $G_{\sF}(y_{(0)},y)=G_{\sF}(y)$.
After performing the Fourier transformation in the time variable,
\begin{equation}
\left\{
\begin{array}{rcl}
     \displaystyle G_{\sF}(y)&=&\displaystyle
     \int \frac{d\omega}{2\pi} e^{-i\omega y^{0}}
     \tilde{G}_{\sF}(\omega,\yy),\\[2mm]
     \displaystyle \delta^{\sN}(y)&=&\displaystyle
     \int \frac{d\omega}{2\pi} e^{-i\omega y^{0}}\delta^{\sN-1}(\yy),
\end{array}
\right.
\label{fou:GF}
\end{equation}
We obtain from Eq.(\ref{eq:GF}),
\begin{equation}
     \left(\Box_{\sN-1}-\xi R-(m^{2}+\omega^{2})\right)
     \tilde{G}_{\sF}(\omega,\yy)
     =-\frac{1}{\sqrt{g}}\delta^{\sN-1}(\yy)\, ,
\label{eq:tildeGF}
\end{equation}
where $\yy$ denotes the space component of $y$.
Equation (\ref{eq:tildeGF}) has the same form as the one for scalar
Green functions on $S^{\sN-1}$.
For the maximally symmetric space $S^{\sN-1}$ we can easily solve
the equation for the Green function following the method developed by
Allen and Jacobson.\cite{AJ}
The Green function $\tilde{G}_{\sF}(\omega,\yy)$ is obtained
straightforwardly.
Because of the symmetry on $R \otimes S^{\sN-1}$,
$\tilde{G}_{\sF}(\omega,\yy)$ is represented as a function only of
$\sigma = a\theta$ which is
the geodesic distance between $\yy_{(0)}$ and $\yy$ on $S^{\sN-1}$.
The Laplacian acting on a function of $\sigma$ is found to be \cite{AJ}
\begin{eqnarray}
     \Box_{\sN-1}f(\sigma)
     =\frac{1}{a^{2}}(\sin\theta)^{2-\sN}\frac{d}{d\theta}(\sin\theta)^{\sN-2}
     \frac{d}{d\theta}\ f(\sigma)\nonumber \\
     =\left({\partial_{\sigma}}^{2}
     +\frac{N-2}{a}\cot\left(\frac{\sigma}{a}\right)
      \partial_{\sigma}\right)\ f(\sigma)\, .
\label{lap:g}
\end{eqnarray}
Using Eq.(\ref{lap:g}) we rewrite the Eq.(\ref{eq:tildeGF}) to obtain
\begin{equation}
     \left({\partial_{\sigma}}^{2}
          +\frac{N-2}{a}\cot\left(\frac{\sigma}{a}\right)\partial_{\sigma}
          -\xi R -(m^{2}+\omega^{2})
     \right)\tilde{G}_{\sF}=0\, ,
\label{eq:sigGF}
\end{equation}
where we restrict ourselves to the region  $\sigma\neq 0$.
To solve Eq.(\ref{eq:sigGF}) we make a change of the variable
$\displaystyle z=\cos^{2}\left(\frac{\sigma}{2a}\right)$ and find
\begin{equation}
     \left[z(1-z){\partial_{z}}^{2}
           +\left(\frac{N-1}{2}-(N-1)z\right)\partial_{z}
           -(N-1)(N-2)\xi-(m^{2}+\omega^{2})a^{2}
     \right]\tilde{G}_{\sF}=0\, .
\label{eq:HG}
\end{equation}
Equation (\ref{eq:HG}) is known as the hypergeometric differential equation.
The solution of this equation is given by the linear combination of
hypergeometric functions,\cite{MF}
\begin{eqnarray}
     \tilde{G}_{\sF}
     &=&q F\left(\frac{N-2}{2}+i\alpha, \frac{N-2}{2}-i\alpha,
             \frac{N-1}{2};z\right)\nonumber \\
     &&+p F\left(\frac{N-2}{2}+i\alpha, \frac{N-2}{2}-i\alpha,
             \frac{N-1}{2};1-z\right)\, ,
\label{sol:GF}
\end{eqnarray}
where we define $\alpha$ by
\begin{equation}
     \alpha = \sqrt{(m^{2}+\omega^{2})a^{2}+(N-1)(N-2)\xi
             -\frac{(N-2)^{2}}{4}}\, .
\end{equation}
As Green function $\tilde{G}_{\sF}$ is regular at the point $\sigma=a\pi$,
we find that $p=0$.
To determine the constant $q$ we consider the singularity \cite{MF}
of $\tilde{G}_{\sF}$ in the limit $\sigma\rightarrow 0$
\begin{equation}
     \tilde{G}_{\sF}\longrightarrow q
     \frac{\displaystyle \Gamma\left(\frac{N-1}{2}\right)
                         \Gamma\left(\frac{N-3}{2}\right)}
          {\displaystyle \Gamma\left(\frac{N-2}{2}+i\alpha\right)
                         \Gamma\left(\frac{N-2}{2}-i\alpha\right)}
     \left(\frac{\sigma}{2a}\right)^{3-\sN}\, ,
\label{lim:gf}
\end{equation}
and compare it with the singularity of the Green function
in flat space-time.
This procedure is justified because the singularity on a curved space-time
background has the same structure as that in the flat space-time.
For $\sigma \sim 0$ the Green function in the flat space-time behaves as
\begin{equation}
     {\tilde{G}_{\sF}}^{flat}(\sigma)\sim\frac{1}{4\pi^{(\sN-1)/2}}
                        \Gamma\left(\frac{N-3}{2}\right)
                        \sigma^{3-\sN}\, .
\label{fla:g}
\end{equation}
Comparing Eq.(\ref{lim:gf}) with Eq.(\ref{fla:g}) we obtain the constant
$q$ :
\begin{equation}
     q=\frac{a^{3-\sN}}{(4\pi)^{(\sN-1)/2}}
     \frac{\displaystyle \Gamma\left(\frac{N-2}{2}+i\alpha\right)
                         \Gamma\left(\frac{N-2}{2}-i\alpha\right)}
          {\displaystyle \Gamma\left(\frac{N-1}{2}\right)}\, .
\label{const:q}
\end{equation}
{}From Eqs.(\ref{fou:GF}),(\ref{sol:GF}) and (\ref{const:q})
we finally obtain the scalar two-point function $G_{\sF}$,
\begin{equation}
\begin{array}{l}
     \displaystyle G_{\sF}(y)=\frac{a^{3-\sN}}{(4\pi)^{(\sN-1)/2}}
     \int \frac{d\omega}{2\pi} e^{-i\omega y^{0}}
     \frac{\displaystyle \Gamma\left(\frac{N-2}{2}+i\alpha\right)
                         \Gamma\left(\frac{N-2}{2}-i\alpha\right)}
          {\displaystyle \Gamma\left(\frac{N-1}{2}\right)}\\
     \displaystyle\times
     F\left(\frac{N-2}{2}+i\alpha, \frac{N-2}{2}-i\alpha,
             \frac{N-1}{2};\cos^{2}\left(\frac{\sigma}{2a}\right)\right)\, .
\label{sctf}
\end{array}
\end{equation}
The two-point function (\ref{sctf}) develops many singularities at
$\sigma=2\pi n a$ where $n$ is an arbitrary integer.
This property is a direct consequence of the boundedness of the
space $S^{\sN-1}$.
In other words the geodesic distance $\sigma$ is bounded in $[0,2\pi a)$.
Thus the two-point function (\ref{sctf}) satisfies the periodic boundary
condition $G_{\sF}(y^{0},\sigma)=G_{\sF}(y^{0},\sigma+2\pi n a)$.
In the two dimensional limit $N \rightarrow 2$
the two-point function reduces to the well-known formula
\begin{eqnarray}
     G_{\sF}(y) &\rightarrow& \frac{a}{2}
     \int \frac{d\omega}{2\pi} e^{-i\omega y^{0}}
     \frac{\cosh (\alpha(\pi-\sigma /a))}{\alpha \sinh (\alpha\pi)}
     \nonumber \\
     &=&\frac{1}{2\pi a} \sum_{n=-\infty}^{\infty}
     \int \frac{d\omega}{2\pi} e^{-i p y}\frac{1}{p^{2}+m^{2}} \, ,
\label{g:2d}
\end{eqnarray}
where $p^{\mu}$ and $y^{\mu}$ are given by
\begin{equation}
\left\{
\begin{array}{rcl}
     p^{\mu}&=&\displaystyle\left(\omega,\frac{n}{a^{2}}\right) \, , \\
     y^{\mu}&=&(y^{0},\theta)
     \, =\, \displaystyle\left(y^{0},\frac{\sigma}{a}\right)\, .
\end{array}
\right.
\label{py:2d}
\end{equation}
In deriving Eq.(\ref{g:2d}) we employ the following summation formula,\cite{MF}
\begin{equation}
     \frac{1}{\pi}\sum_{n=-\infty}^{\infty}\frac{\cos (nx)}{n^{2}+a^{2}}
     =\frac{\cosh (a(\pi-x))}{\sinh (a\pi)}\, .
\end{equation}
Equation (\ref{g:2d}) is nothing but the ordinary two-point function
with the periodic boundary condition for the spatial coordinate.

If $m=0$ and $\displaystyle \xi=\frac{N-2}{4(N-1)}$, the field equation
(\ref{eq:GF}) is invariant under conformal transformation. In this case
we can perform the Fourier integral in Eq.(\ref{sctf}) at $y=y_{(0)}$,
\begin{equation}
     G_{\sF}^{conformal}(y=y_{(0)})
     =\frac{2a^{2-\sN}}{(4\pi)^{(\sN+1)/2}}
     \Gamma\left(\frac{N-1}{2}\right)\Gamma\left(\frac{N}{2}-1\right)
     \Gamma\left(1-\frac{N}{2}\right)\, .
\label{int:GF0}
\end{equation}
The two-point function at $y=y_{(0)}$ obtained above is known to be useful
in calculating the effective
potential and studying the vacuum structure of the theory.\cite{CR}

Next we consider the spinor two-point function on $R \otimes S^{\sN-1}$.
The spinor two-point function $D$ is defined by the Dirac equation
\begin{equation}
     (\nab +m)D(y)=-\frac{1}{\sqrt{g}}\delta^{\sN}(y)\, .
\label{eq:D}
\end{equation}
We introduce the bispinor function $G$ defined by
\begin{equation}
     (\nab -m)G(y)=D(y)\, .
\label{DfromG}
\end{equation}
According to Eq.(\ref{eq:D}) $G(y)$ satisfies the following equation,
\begin{equation}
     (\nab\nab -m^{2})G(y)=-\frac{1}{\sqrt{g}}\delta^{\sN}(y)\, .
\label{eq:Gsp}
\end{equation}
On $R \otimes S^{\sN-1}$ we rewrite Eq.(\ref{eq:Gsp}) in the following
form
\begin{equation}
     \left((\partial_{0})^{2}+\Box_{\sN-1}-\frac{R}{4}-m^{2}\right)G(y)
     =-\frac{1}{\sqrt{g}}\delta^{\sN}(y)\, .
\label{eq:G}
\end{equation}
where $\Box_{\sN-1}$ is the Laplacian on $S^{\sN-1}$.
Performing the Fourier transformation
\begin{equation}
     G(y)=\int \frac{d\omega}{2\pi} e^{-i\omega y^{0}}\tilde{G}(\omega,\yy)\, ,
\label{fou}
\end{equation}
we rewrite Eq.(\ref{eq:G}) in the form
\begin{equation}
     \left(\Box_{\sN-1}-\frac{R}{4}-(m^{2}+\omega^{2})\right)
     \tilde{G}(\omega,\yy)
     =-\frac{1}{\sqrt{g}}\delta^{\sN-1}(\yy)\, .
\label{eq:tildeG}
\end{equation}
Equation (\ref{eq:tildeG}) is of the same form as the one for the spinor
Green function on $S^{\sN-1}$.
According to the method developed by Camporesi,\cite{CA} we can find the
expression of $\tilde{G}(\omega,\yy)$.
The general form of the Green function $\tilde{G}(\omega,\yy)$
is written as
\begin{equation}
     \tilde{G}(\omega,\yy)=U(\yy)(g_{\sN-1}(\omega,\sigma)
     +g^{\prime}_{\sN-1}(\omega,\sigma)n_{i}\gamma^{i})\, ,
\label{ansatz}
\end{equation}
where $U$ is a matrix in the spinor indices,
$g_{\sN-1}$ and $g^{\prime}_{\sN-1}$ are scalar functions only of
$\omega$ and $\sigma$, $n_{i}$ is a unit vector tangent to the geodesic
$n_{i}=\nabla_{i}\sigma$, $\gamma^{i}$ is the Dirac gamma matrices and
Roman index $i$ runs over the space components $(i=1,2,\cdots,N-1)$.
Inserting Eqs.(\ref{ansatz}) into Eq.(\ref{eq:tildeG}) we get
\begin{equation}
\begin{array}{l}
     \displaystyle
     \Biggl[U\Box_{\sN-1}(g_{\sN-1}+g^{\prime}_{\sN-1}n_{i}\gamma^{i})
     +2(\nabla_{j}U)\nabla^{j}
     (g_{\sN-1}+g^{\prime}_{\sN-1}n_{i}\gamma^{i}) \\
     \displaystyle
     +(\Box_{\sN-1}U)(g_{\sN-1}+g^{\prime}_{\sN-1}n_{i}\gamma^{i})
     -\left(\frac{R}{4}+(m^{2}+\omega^{2})\right)U
     (g_{\sN-1}+g^{\prime}_{\sN-1}n_{i}\gamma^{i})\Biggr]
     =0\, ,
\end{array}
\label{eq:sn}
\end{equation}
where we restrict ourselves to the region $\sigma\neq 0$.
To evaluate Eq.(\ref{eq:sn}) we have to calculate the
covariant derivative of $U$ and $n_{i}$.
U is the operator which makes parallel transport of the spinor at
point $\yy_{(0)}$
along the geodesic to point $\yy$.
Thus the operator $U$ must satisfy the following parallel transport
equations:\cite{AL}
\begin{equation}
\left\{
\begin{array}{rcl}
     n^{i}\nabla_{i}U&=&0\, ,\\
     U(\yy_{(0)})&=&\11\, ,
\end{array}
\right.
\label{eq:para}
\end{equation}
Here we set
\begin{equation}
     \nabla_{i}U \equiv V_{i}U\, .
\end{equation}
{}From the integrability condition \cite{CR} on $V_{i}$,
\begin{equation}
     \nabla_{i}V_{j}-\nabla_{j}V_{i}-[V_{i},V_{j}]
     =\frac{1}{a^{2}}\sigma_{ij}\, ,
\label{cond:int}
\end{equation}
and the parallel transport equation (\ref{eq:para}) we easily find that
\begin{equation}
     V_{i}=-\frac{1}{a}\tan\left(\frac{\sigma}{2a}\right)\sigma_{ij}n^{j}\, ,
\end{equation}
where $\sigma_{\mu\nu}$ are the antisymmetric tensors constructed by the Dirac
gamma matrices,
$\displaystyle \sigma_{\mu\nu}=\frac{1}{4}[\gamma_{\mu},\gamma_{\nu}]$.
To find $V_{i}$ we have used the fact that the maximally symmetric bitensors
on $S^{\sN-1}$ is represented as a sum of products of $n_{i}$ and
$g_{ij}$ with coefficients which are functions only of $\sigma$.
After some calculations we get the Laplacian acting on $U$
\begin{equation}
     \Box_{\sN-1}U=-\frac{N-2}{4a^{2}}
     \tan^{2}\left(\frac{\sigma}{2a}\right)U\, .
\label{lap:U}
\end{equation}
The first and second derivatives of $n_{i}$ are also the maximally symmetric
bitensors and found to be \cite{AJ}
\begin{equation}
\begin{array}{rcl}
     \nabla_{i}n_{j}&=&\displaystyle
     \frac{1}{a}\cot\left(\frac{\sigma}{a}\right)
     (g_{ij}-n_{i}n_{j})\, ,\\
     \Box_{\sN-1}n_{i}&=&\displaystyle -\frac{1}{a^{2}}\cot^{2}
     \left(\frac{\sigma}{a}\right)(N-2)n_{i}\, .
\end{array}
\label{der:n}
\end{equation}
Therefore Eq.(\ref{eq:sn}) reads
\begin{equation}
\begin{array}{l}
     \displaystyle
     \left({\partial_{\sigma}}^{2}
          +\frac{N-2}{a}\cot\left(\frac{\sigma}{a}\right)\partial_{\sigma}
          -\frac{N-2}{4a^{2}}\tan^{2}\left(\frac{\sigma}{2a}\right)
          -\frac{R}{4}-(m^{2}+\omega^{2})
     \right)g_{\sN-1}  \\
     \displaystyle
     +n_{i}\gamma^{i}\left({\partial_{\sigma}}^{2}
          +\frac{N-2}{a}\cot\left(\frac{\sigma}{a}\right)\partial_{\sigma}
          -\frac{N-2}{4a^{2}}\cot^{2}\left(\frac{\sigma}{2a}\right)
          -\frac{R}{4}-(m^{2}+\omega^{2})
     \right)g^{\prime}_{\sN-1}=0\, .
\end{array}
\label{eq:sig}
\end{equation}
Since the two terms in Eq.(\ref{eq:sig})
are independent, each term in the left-hand side of Eq.(\ref{eq:sig})
has to vanish.
We define the functions $h_{\sN-1}(\omega,\sigma)$ and
$h^{\prime}_{\sN-1}(\omega,\sigma)$
by $\displaystyle g_{\sN-1}(\omega,\sigma)$ $
=\cos\left(\frac{\sigma}{2a}\right)h_{\sN-1}(\omega,\sigma)$ and
$\displaystyle g^{\prime}_{\sN-1}(\omega,\sigma)$ $
=\sin\left(\frac{\sigma}{2a}\right)h^{\prime}_{\sN-1}(\omega,\sigma)$
respectively and make a change of variable by
$\displaystyle z=\cos^{2}\left(\frac{\sigma}{2a}\right)$.
We then find that Eq.(\ref{eq:sig}) is rewritten in the form of
two hypergeometric differential equations :
\begin{equation}
     \left[z(1-z){\partial_{z}}^{2}
           +\left(\frac{N+1}{2}-Nz\right)\partial_{z}
           -\frac{(N-1)^{2}}{4}-(m^{2}+\omega^{2})a^{2}
     \right]h_{\sN-1}(\omega,z)=0\, .
\label{eq:HGsp}
\end{equation}
\begin{equation}
     \left[z(1-z){\partial_{z}}^{2}
           +\left(\frac{N-1}{2}-Nz\right)\partial_{z}
           -\frac{(N-1)^{2}}{4}-(m^{2}+\omega^{2})a^{2}
     \right]h^{\prime}_{\sN-1}(\omega,z)=0\, .
\label{eq:HGspp}
\end{equation}
Noting that the Green function is regular at the point $\sigma=a\pi$
we write the solutions of Eqs.(\ref{eq:HGsp}) and (\ref{eq:HGspp})
by only one kind of the hypergeometric function,\cite{MF}
\begin{equation}
     h_{\sN-1}(\omega,z)
     =c_{\sN-1}F\left(\frac{N-1}{2}+i\beta, \frac{N-1}{2}-i\beta,
             \frac{N+1}{2};z\right)\, ,
\label{eq:h}
\end{equation}
\begin{equation}
     h^{\prime}_{\sN-1}(\omega,z)
     =c^{\prime}_{\sN-1}F\left(\frac{N-1}{2}+i\beta, \frac{N-1}{2}-i\beta,
             \frac{N-1}{2};z\right)\, ,
\label{eq:hp}
\end{equation}
where we define $\beta$ by
\begin{equation}
     \beta = a\sqrt{m^{2}+\omega^{2}}\, .
\end{equation}
As we remained in the region where $\sigma\neq 0$ the normalization
constants $c_{\sN-1}$ and $c^{\prime}_{\sN-1}$ are yet undetermined.
To obtain $c_{\sN-1}$ and $c^{\prime}_{\sN-1}$ we consider
the singularity \cite{MF} of ${\tilde{G}}$ in the limit
$\sigma \rightarrow 0$,
\begin{eqnarray}
     \tilde{G}&\longrightarrow& c_{\sN-1}
     \frac{\displaystyle \Gamma\left(\frac{N+1}{2}\right)
                         \Gamma\left(\frac{N-3}{2}\right)}
          {\displaystyle \Gamma\left(\frac{N-1}{2}+i\beta\right)
                         \Gamma\left(\frac{N-1}{2}-i\beta\right)}
     \left(\frac{\sigma}{2a}\right)^{3-\sN}
     \nonumber \\
     &&+n_{i}\gamma^{i}c^{\prime}_{\sN-1}
     \frac{\displaystyle \left(\Gamma\left(\frac{N-1}{2}\right)\right)^{2}}
          {\displaystyle \Gamma\left(\frac{N-1}{2}+i\beta\right)
                         \Gamma\left(\frac{N-1}{2}-i\beta\right)}
     \left(\frac{\sigma}{2a}\right)^{3-\sN}\, ,
\label{lim:g}
\end{eqnarray}
Comparing Eq.(\ref{lim:g}) with Eq.(\ref{fla:g}), the over-all factors
$c_{\sN-1}$ and $c^{\prime}_{\sN-1}$ are obtained :
\begin{eqnarray}
     c_{\sN-1}&=&\frac{a^{3-\sN}}{(4\pi)^{(\sN-1)/2}}
     \frac{\displaystyle \Gamma\left(\frac{N-1}{2}+i\beta\right)
                         \Gamma\left(\frac{N-1}{2}-i\beta\right)}
          {\displaystyle \Gamma\left(\frac{N-1}{2}\right)}\, ,
\label{const} \\
     c^{\prime}_{\sN-1}&=&0\, .
\label{constp}
\end{eqnarray}
Therefore $g^{\prime}_{\sN-1}(\omega,\sigma)$ disappears in the final
expression of the Green function.
{}From Eqs.(\ref{eq:h}) and (\ref{const})
we find the expression of $\tilde{G}(\omega,\yy)$,
\begin{equation}
\begin{array}{l}
    \displaystyle \tilde{G}(\omega,\yy)=U(\yy)g_{\sN-1}(\omega,\sigma) \\
    =\displaystyle U(\yy)\frac{a^{3-\sN}}{(4\pi)^{(\sN-1)/2}}
     \frac{\displaystyle\Gamma\left(\frac{N-1}{2}+i\beta\right)
                        \Gamma\left(\frac{N-1}{2}-i\beta\right)}
          {\displaystyle\Gamma\left(\frac{N+1}{2}\right)}\\[2mm]
    \displaystyle \times\cos\left(\frac{\sigma}{2a}\right)
      F\left(\frac{N-1}{2}+i\beta, \frac{N-1}{2}-i\beta,
             \frac{N+1}{2};\cos^{2}\left(\frac{\sigma}{2a}\right)
       \right)\, .
\end{array}
\label{sol:g}
\end{equation}
Thus the Green function $G(y)$ on $R\otimes S^{\sN-1}$ is obtained.

The spinor two-point function $D(y)$ is derived from the
Green function $G(y)$.
Inserting Eqs.(\ref{fou}) and (\ref{ansatz}) into Eq.(\ref{DfromG})
we get
\begin{eqnarray}
     D&=&\int\frac{d\omega}{2\pi}e^{-i\omega y^{0}}
     (-i\omega\gamma^{0}+\gamma^{i}\nabla_{i}-m) U g_{\sN-1}
     \nonumber \\
     &=&\int\frac{d\omega}{2\pi}e^{-i\omega y^{0}}
     \left[\gamma_{i}n^{i}U\left(\partial_{\sigma}
           -\frac{N-2}{2a}\tan\left(\frac{\sigma}{2a}\right)\right)
           g_{\sN-1}-(i\omega\gamma^{0}+m) U g_{\sN-1}\right]\, .
\label{eq:D:fin}
\end{eqnarray}
Substituting Eq.(\ref{sol:g}) in Eq.(\ref{eq:D:fin}) the spinor
two-point function $D(y)$ is obtained
\begin{equation}
\begin{array}{l}
    \displaystyle D(y)=\displaystyle-\frac{a^{3-\sN}}{(4\pi)^{(\sN-1)/2}}
    \int\frac{d\omega}{2\pi}e^{-i\omega y^{0}}
     \frac{\displaystyle\Gamma\left(\frac{N-1}{2}+i\beta\right)
                        \Gamma\left(\frac{N-1}{2}-i\beta\right)}
          {\displaystyle\Gamma\left(\frac{N+1}{2}\right)}\\[2mm]
    \displaystyle
       \times\biggl[(i\omega\gamma^{0}+m) U(\yy)
       \cos\left(\frac{\sigma}{2a}\right)
      F\left(\frac{N-1}{2}+i\beta, \frac{N-1}{2}-i\beta,
             \frac{N+1}{2};\cos^{2}\left(\frac{\sigma}{2a}\right)
       \right)\\[2mm]
    \displaystyle
       +\gamma_{i}n^{i}U(\yy)\frac{N-1}{2a}\sin\left(\frac{\sigma}{2a}\right)
      F\left(\frac{N-1}{2}+i\beta, \frac{N-1}{2}-i\beta,
             \frac{N-1}{2};\cos^{2}\left(\frac{\sigma}{2a}\right)
       \right)\biggr]\, .
\end{array}
\label{stf}
\end{equation}
According to the anticommutation relation of spinor fields
the two-pint function (\ref{stf}) satisfies the antiperiodic boundary
condition $D(y^{0},\sigma)=-D(y^{0},\sigma +2 \pi n a)$
where $n$ is an arbitrary integer.
In the two dimensional limit $N \rightarrow 2$ the two-point function
(\ref{stf}) simplifies :
\begin{eqnarray}
     D(y)&\rightarrow&-\frac{a}{2}
     \int\frac{d\omega}{2\pi}e^{-i\omega y^{0}}
     \left[(i\omega\gamma^{0}+m)\frac{\sinh(\beta(\pi-\sigma /a))}
                                   {\beta \cosh(\beta\pi)}
     +a \gamma^{1}\frac{\cosh(\beta(\pi-\sigma /a))}
                     {a \cosh(\beta\pi)}
     \right]                \nonumber \\
     &=&\frac{1}{2\pi a} \sum_{n=-\infty}^{\infty}
     \int\frac{d\omega}{2\pi}e^{-i p y}
     \frac{1}{i\ps-m}\, ,
\label{stpf:2d}
\end{eqnarray}
where $p^{\mu}$ and $y^{\mu}$ are defined by
\begin{equation}
\left\{
\begin{array}{rcl}
     p^{\mu}&=&\displaystyle\left(\omega,\frac{2n-1}{2a^{2}}\right) \, , \\
     y^{\mu}&=&\displaystyle\left(y^{0},\frac{\sigma}{a} \right)\, .
\end{array}
\right.
\label{spy:2d}
\end{equation}
To obtain Eq.(\ref{stpf:2d}) the following formulae were employed,
\begin{equation}
\begin{array}{rcl}
     \displaystyle\frac{1}{\pi}\sum_{n=-\infty}^{\infty}
     \frac{\cos ((2n-1)x)}{(2n-1)^{2}+(2a)^{2}}
     &=&\displaystyle\frac{\sinh (a(\pi-2x))}{4 a \cosh (a\pi)}\, ,\\
     \displaystyle\frac{1}{\pi}\sum_{n=-\infty}^{\infty}
     \frac{(2n-1) \sin ((2n-1)x)}
     {(2n-1)^{2}+(2a)^{2}}
     &=&\displaystyle\frac{\cosh (a(\pi-2x))}{2 \cosh (a\pi)}\, .
\end{array}
\end{equation}
We easily see that the Eq.(\ref{stpf:2d}) is identical to the well-known
two-point function of spinor with antiperiodic boundary condition
for the spatial coordinate.

It should be noted that $\mbox{tr} D(y=y_{(0)})$ is required in evaluating
the effective potential:
\begin{equation}
     \mbox{tr} D(y=y_{(0)})=\displaystyle -
     \frac{\mbox{tr} \11m a^{3-\sN}}{(4\pi)^{(\sN-1)/2}}
     \Gamma\left(\frac{3-N}{2}\right)
     \int\frac{d\omega}{2\pi}
     \frac{\displaystyle \Gamma\left(\frac{N-1}{2}+i\beta\right)
                         \Gamma\left(\frac{N-1}{2}-i\beta\right)}
          {\displaystyle \Gamma\left(1+i\beta\right)
                         \Gamma\left(1-i\beta\right) }\, .
\label{trD}
\end{equation}
For the massless case we can perform the Fourier integral in Eq.(\ref{trD})
and get
\begin{equation}
     \frac{\mbox{tr} D(y=y_{(0)},m=0)}{m}\longrightarrow\displaystyle -
     \frac{2\mbox{tr} \11 a^{2-\sN}}{(4\pi)^{(\sN+1)/2}}
     \Gamma\left(\frac{N}{2}\right)
     \Gamma\left(\frac{N-1}{2}\right)
     \Gamma\left(1-\frac{N}{2}\right)\, .
\label{int:D}
\end{equation}

In the present paper we calculated scalar and spinor two-point functions
on $R\otimes S^{\sN-1}$.
In the final expression for the scalar and spinor two-point functions,
Eqs.(\ref{sctf}) and (\ref{stf}),
the Fourier integral is remained.
For some cases we can perform the Fourier integral in
Eqs.(\ref{sctf}) and (\ref{stf}) at $y=y_{(0)}$.
The resulting explicit expressions may be useful to investigate the vacuum
structure of the Einstein universe.\cite{IIM}

We would like to thank K.~Fukazawa, S.~Mukaigawa and H.~Sato
for helpful discussions.

\end{document}